\documentclass[journal]{IEEEtran}
\usepackage{amsmath,amssymb,amsfonts}

\usepackage[linesnumbered,ruled,vlined]{algorithm2e}
\usepackage{blindtext}
\usepackage{cite}
\usepackage{amsmath,amssymb,amsfonts}
\usepackage[noend]{algpseudocode}
\usepackage{graphicx}
\usepackage{textcomp}
\usepackage{svg}
\usepackage{xcolor}
\usepackage{relsize}

\begin{document}

\title{Deep Reinforcement Learning for Joint Time and Power Management in SWIPT-EH CIoT}

\author{Nadia~Abdolkhani,~\IEEEmembership{Student Member,~IEEE,}
Nada~Abdel~Khalek,~\IEEEmembership{Student Member,~IEEE,}

        Walaa~Hamouda,~\IEEEmembership{Senior Member,~IEEE,}
        and~Iyad~Dayoub,~\IEEEmembership{Senior Member,~IEEE}%
        

\thanks{Manuscript received November xx, 2024; revised xxxx xx, 20xx. This work was supported in part by the Natural Sciences and Engineering Research Council of Canada (NSERC) under Grant 00881, and in part by the Fonds de Recherche du Québec - Nature et Technologies (FRQNT). (Corresponding author: Walaa Hamouda.)}

\thanks{Nadia Abdolkhani, Nada Abdelkhalek, and Walaa Hamouda are with the Department
of Electrical and Computer Engineering,  Concordia University, Montreal,
QC H3G 1M8, Canada (e-mail: n\_abdolk@ece.concordia.ca; n\_abdelk@ece.concordia.ca; hamouda@
ece.concordia.ca).}
\thanks{I. Dayoub is with Université Polytecnique Hauts-de-France, 59313 Valenciennes, France (e-mail: : iyad.dayoub@uphf.fr)}
}


\maketitle

\begin{abstract}
This letter presents a novel deep reinforcement learning (DRL) approach for joint time allocation and power control in a cognitive Internet of Things (CIoT) system with simultaneous wireless information and power transfer (SWIPT). The CIoT transmitter autonomously manages energy harvesting (EH) and transmissions using a learnable time switching factor while optimizing power to enhance throughput and lifetime. The joint optimization is modeled as a Markov decision process under small-scale fading, realistic EH, and interference constraints. We develop a double deep Q-network (DDQN) enhanced with an upper confidence bound. Simulations benchmark our approach, showing superior performance over existing DRL methods.
\end{abstract}


\IEEEpeerreviewmaketitle

\section{Introduction}

\IEEEPARstart{T}{he} rapid proliferation of smartphones, cloud computing, and the Internet of Things (IoT) has significantly intensified the demand for wireless spectrum. This surge has led to heightened interest in leveraging cognitive radio (CR) within IoT networks (CIoT) to dynamically borrow spectrum \cite{nada_survey_2023, deap_learning}. To share the spectrum with primary users (PUs), CIoT devices need to implement a power control strategy to avoid causing interference and maximizing their throughput. This challenge is further exacerbated by the deployment of energy-constrained CIoT devices in remote or hazardous environments, where frequent battery replacements are impractical. Energy harvesting (EH) can provide a sustainable solution to these devices by collecting energy from nearby radio frequency (RF) resources. \textcolor{black}{Unlike wireless power transfer (WPT), simultaneous wireless information and power transfer (SWIPT) allows devices to divide a communication slot between energy harvesting and data transmission using the time-switching (TS) protocol, rather than limit the slot to a single function. Therefore, dynamic decision-making is essential for energy-constrained CIoT devices to intelligently select the optimal TS ratio and transmission power. Deep reinforcement learning (DRL) has been explored to allow CIoT devices to develop \textit{self-adaptation} capabilities and optimize network operations.} However, the application of DRL in CIoT networks is still in its infancy \cite{10673973}.

\subsection{Scope and Contributions of the Work}

Most studies that use DRL to optimize transmission in CRs have primarily focused on either managing power \cite{lu_full_duplex_power_control, 9680720} or time allocation \cite{8885426} separately. Moreover, they overlook the potential of an integrated approach that could significantly enhance data transmission efficiency and EH. While Zhang \textit{et al.} proposed a joint strategy to optimize both time and power, they adopted a multi-agent D2D perspective, where CRs not only share common objectives but can also harvest energy from each other's transmissions \cite{9915473}. Achieving common objectives often requires significant communication overhead, which can render it impractical for energy-constrained CIoT. Additionally, this approach may encounter convergence issues. The assumption that all CRs have aligned goals is often unrealistic, as often times they may have competing interests. Similarly, Omidkar \textit{et al.} considered the joint optimization of time and power allocation; however, they rely on offline optimization approaches which require prior knowledge and assumptions about the radio environment \cite{R2}. Tashman \textit{et al.} developed a DRL-based power control strategy for SWIPT-EH CR networks using the TS protocol, however the TS factor is treated as a fixed hyperparameter \cite{tashma_drl}. This approach not only demands prior knowledge for optimal performance, but limits its adaptability to environment changes. Abdel Khalek \textit{et al.} investigated the selection of optimal operation modes in WPT-enabled CIoT, where devices can transmit or harvest energy within a communication slot.

Given the aforementioned gap, we introduce a novel DRL-based approach to jointly optimize time allocation and transmit power, designed to maximize both long-term throughput and lifetime for a SWIPT-EH CIoT network. This approach treats both the SWIPT TS factor and the transmit power as learnable parameters by the CIoT agent. We also adopt a realistic EH scenario where the CIoT agent recharges from ambient sources without a stable dedicated source.  We formulate the DRL framework as a discrete-time model-free Markov decision process (MDP) with continuous states and discrete actions, and a novel lightweight double deep Q-network (DDQN) architecture is proposed. This DDQN is designed to autonomously learn an operation strategy that does not require any prior knowledge, while considering factors such as channel occupancy, fading conditions, energy arrival patterns, and interference constraints. Additionally, upper confidence bound (UCB) exploration is utilized to optimize dynamic decision-making. 

\section{System Model and Problem Formulation}

Consider an CIoT network consisting of a transmitter-receiver (Tx-Rx) pair that shares the spectrum alongside a primary Tx-Rx pair. Similar to \cite{IoT_2024_Nada_Nadia}, the communication system is divided into time slots, each lasting \(\tau\) seconds, with a total of \( T \) equal-duration slots. The CIoT Tx has a finite battery capacity $B_{max}$ and is capable of SWIPT-EH. The time switching (TS) protocol is employed to manage the SWIPT operation in each time slot $t$ based on the TS factor $0 \leq \rho_t \leq 1$. That is, the CIoT agent is capable of alternating between EH and data transmission, where the duration of each time slot $\tau$ is divided into two periods, one for EH $\rho_t \tau$, and the other for data transmission $(1 - \rho_t) \tau$.

The number of time slots allocated for transmissions by the PU Tx is \(L\), and the PU Tx is capable of maintaining a consistent transmit power of \(P^t_p\) across \(L\) slots. If the PU Tx is active during slot $t$, the PU status indicator \(\omega_p^t\) is set to 1; otherwise, it is 0. In underlay CR, the CIoT Tx is permitted to utilize the same slot as the PU Tx, provided it complies with the interference threshold \(I_{th}\). This threshold is defined by \(P_{s}^t g_{sp}^t \leq I_{th}\), where \(P_{s}^t\) is the transmit power of the CIoT Tx and \(g_{sp}^t\) is the channel power gain between the CIoT Tx and the PU Tx. We model the channel power gains as independently and identically distributed (i.i.d) Rayleigh fading channels for the CIoT Tx-Rx pair $g_{ss}^t$, the PU Tx and CIoT Rx $g_{ps}^t$, and the CIoT Tx and PU Rx $g_{sp}^t$. These channel gains are assumed to stay constant for the duration of each time slot \cite{IoT_2024_Nada_Nadia}. The channel power gain $g_{ij}^t$ follows an exponential distribution with the following probability density function $f_{g_{ij}^t}\left(y\right)=\lambda_{ij}\exp{\left(\lambda_{ij}y\right)}$. The channel fading parameter $\lambda_{ij}$ is influenced by the distance between devices $d_{ij}$ and the path-loss exponent $\alpha$. Specifically, $\lambda_{ij} = d_{ij}^{-\alpha}$.

When the channel is unoccupied, the achievable rate of the CIoT Tx during the \(t\)-th time slot is

\begin{equation}
 R_0^t = \log_2\bigg(1+\frac{P_s^t g_{ss}^t}{\sigma^2}\bigg),
\end{equation}\normalsize
where \(\sigma^2\) denotes the variance of the channel noise. If the PU Tx occupies the channel during the \(t\)-th time slot, the achievable rate of the CIoT Tx is reduced due to interference from the PU, as specified by

\begin{equation}
 R_1^t = \log_2\bigg(1+\frac{P_s^t g_{ss}^t}{P^t_p g_{ps}^t + \sigma^2}\bigg).
\end{equation}\normalsize

\textcolor{black}{The EH process is modeled as an energy arrival model characterized by independent and identically distributed time slots. The energy harvested at $t=0$ is $e_0=0$. We assume that the energy of ambient sources follows a Gamma distribution $\hat{e} \sim \Gamma(k,\beta)$ with shape $k$ and scale $\beta$. Consequently, the energy harvested during each time slot is $e_t = \mu \hat{e}$, where $0 \leq \mu \leq 1$ is the energy conversion efficiency factor. In the next time slot $t+1$, the available battery is updated based on the CIoT device's decision ($\rho_t$, $P_s^t$) as
\smaller$B_{t+1} = \text{min}\big\{
    B_t + \rho_t e_t\tau -(1-\rho_t)P^t_s\tau, B_{max}\big\}$.\normalsize~$\rho_t$ indicates the fraction of the time slot's duration that the CIoT device decided to harvest and (1-$\rho_t$) indicates the remaining time slot's duration that the CIoT will transmit data.}

In the studied CR network, the CIoT Tx aims to optimize its transmission rate while accounting for the interference threshold \(I_{th}\), the available battery energy \(B_t\), and the energy harvested \(e_t\). Thus, the maximization of the CIoT Tx's rate can be formulated as a constrained optimization problem as follows

\begin{subequations}
\label{eqn:optim}
\begin{align}
   &\max_{\rho_t,P^t_s} \sum_{t=1}^{T} (1-\rho_t)\tau\big{[}(1-\omega_{p}^t)R^t_0+\omega_{p}^tR^t_1\big{]} \label{maximization}\\
   &\text{s.t.  } \sum_{t=1}^{T} P^t_s(1-\rho_t)\tau \leq B_0+\sum_{t=0}^{T-1}e_t\tau, ~~\forall T  \label{constraint1}\\
   & ~~~~~0\leq P^t_s(1-\rho_t)\tau\leq B_t,~~ \rho_t\in[0,1] \label{constraint2}\\
   & ~~~~~\omega_{p}^tg_{sp}^tP^t_s\leq I^t_{th},~~ \omega_{p}^t\in\Omega\triangleq\{0,1\}. \label{constraint4}
\end{align}
\end{subequations}\normalsize

\section{Deep Reinforcement Learning for Joint Time and Power Management}

\subsection{The Model-Free Markov Decision Process}

\begin{figure}
    \centering
    \includegraphics[width=0.97\columnwidth]{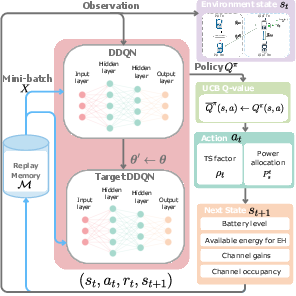}
    \caption{The proposed DDQN-UCB algorithm}
    \label{fig:dqn_model}
\end{figure}

We address the optimization problem by modeling the CIoT Tx action-taking process as a stochastic control process in a discrete-time system. Per the Markov property, the system's next state depends only on its current state and action. In our model, the stochastic behavior of states, like PU activities and energy arrival, adheres to the Markov property. Thus, learning the optimal operation strategy can be described as a Markov decision process (MDP). This MDP is characterized by the tuple \((\boldsymbol{\mathcal{S}},\boldsymbol{\mathcal{A}},\boldsymbol{\mathcal{P}},\boldsymbol{\mathcal{R}},T)\), where \(\boldsymbol{\mathcal{S}}\) represents the set of states within the environment, \(\boldsymbol{\mathcal{A}}\) represents the set of possible actions, \(\boldsymbol{\mathcal{P}}\) is the set of probabilities for transitioning between states, \(\boldsymbol{\mathcal{R}}\) denotes the rewards for each state-action pair, and \(T\) represents the time step. \textcolor{black}{In practical CIoT scenarios, it is difficult to precisely determine the probability density function of energy and channel fading. Moreover, the CIoT network lacks knowledge of the state transition probabilities for each occupancy state of the primary network, making it nearly impossible to accurately determine \(\boldsymbol{\mathcal{P}}\). Consequently, we utilize a model-free MDP approach and employ a deep reinforcement learning (DRL) framework to approximate \(\boldsymbol{\mathcal{R}}\) based on \(\boldsymbol{\mathcal{S}}\) and \(\boldsymbol{\mathcal{A}}\), without relying on \(\boldsymbol{\mathcal{P}}\).} Thus, the model-free MDP tuple is represented as \((\boldsymbol{\mathcal{S}},\boldsymbol{\mathcal{A}},\boldsymbol{\mathcal{R}},T)\) with the following components:

\begin{enumerate}
    \item \textbf{State Space $\boldsymbol{\mathcal{S}}$:} The state space encompasses all possible states across \(T\) time slots. For each state \(s_t\) within the environment, the CIoT agent must consider several factors: the current battery level \(B_t\), the energy harvested in the preceding time slot \(e_{t-1}\), the occupancy status of the slot by the PU Tx, and the channel power gains \(g_{ps}^t\), \(g_{sp}^t\), \(g_{ss}^t\). Thus, the state of the environment at any given time slot \(t\) is $ s_t = \{ B_t, e_{t-1}, \omega_{p}^t, g_{ps}^t, g_{sp}^t, g_{ss}^t\}$.
    \item \textbf{Action Space $\boldsymbol{\mathcal{A}}$:} The action space consists of all possible actions available to the CIoT agent. Based on the current state of the environment \(s_t\), the CIoT agent must decide on the value of the TS factor $\rho_t$. Additionally, the agent must determine the appropriate transmit power \(P_s^t\). Thus, the action taken by the CIoT agent at each time slot \(t\) is $ a_t = [\rho_t, P_s^t]$, where \(\rho_t \in [0,1]\) and \(P_s^t \in P\).
    \item \textbf{Reward $\boldsymbol{\mathcal{R}}$:} The reward value is determined by the achievable rate, ensuring that all constraints specified in (\ref{eqn:optim}) are met. Any action \(a_t\) by the CIoT agent that violates the constraints in (\ref{eqn:optim}) results in a negative reward (penalty). Consequently, the reward \(r_t\) for the CIoT at each time slot \(t\) is:
    
\begin{equation}\label{eq:reward}
 r_t = \begin{cases} \rho_t'R_0^t & \omega_{p}^t=0,0\leq P_s^t\rho_t'\tau \leq B_t\\
 \rho_t'R_1^t & \omega_{p}^t=1,0\leq P_s^t\rho_t'\tau\leq B_t, P_s^t g_{sp}^t\leq I_{th}\\
 -\phi & \text{others,} 
\end{cases}
\end{equation}\normalsize
where $\rho_t'$ represents $(1-\rho_t)$.
\item \textbf{Time Step $T$:} We characterize each progression from time slot \(t\) to \(t+1\) as a single step and systematically evaluate each state-action across all time slots \(T\).
\end{enumerate}
Fig.~\ref{fig:dqn_model} provides a detailed view of the employed DRL algorithm, including our proposed UCB exploration strategy, which will be discussed in the next subsections.

\subsection{Proposed DDQN-UCB Strategy}
In this subsection, we propose a DRL framework that enables the CIoT agent to approximate the state-value function and learn a strategy \(\pi\) for selecting actions based on the current environment state. By employing \(\pi\), the agent aims to optimize the long-term cumulative reward (rate) while conforming to the constraints of the CIoT system. \textcolor{black}{Since energy-constrained CIoT devices operate at low transmission powers and the time-switching (TS) ratio is bounded by \( 0 \leq \rho \leq 1 \), the action space is relatively small, allowing for efficient discretization to reduce computational complexity. Deep Q-networks (DQNs) have been notoriously used to learn optimal actions in discrete action spaces, however, they suffer from overestimation bias that could lead the learning agent to choose over-optimistic actions, leading to suboptimal performance. To overcome this, we propose a double deep Q-network (DDQN) to predict the cumulative reward (Q-value) for a possible action \(a\) in a given state \(s\).} Essentially, the DDQN optimizes its parameters \(\boldsymbol{\theta}\) to ensure that \(Q^\pi(s,a;\boldsymbol{\theta}) \approx Q^\pi(s,a)\). \textcolor{black}{The DDQN is implemented as a lightweight fully connected neural network, where the input layer contains \(j\) neurons reflecting the dimensions of the state space \(\boldsymbol{\mathcal{S}}\). It also includes two hidden layers with \(h_1\) and \(h_2\) neurons respectively, and an output layer with \(z\) neurons. }The DDQN parameters $\boldsymbol{\theta}= \{\textbf{W}^{(i)}, \textbf{b}^{(i)}\}$ for each network layer $i$, represent the weights and biases.

During training, a Target DDQN is utilized, which initially mirrors the DDQN. As training advances, the parameters of the Target DDQN are updated at a slower rate compared to those of the DDQN, often across several training steps. The proposed DDQN is trained using the mean squared error (MSE) loss \(\mathcal{L}\) to compute the MSE between predicted and target Q-values for a mini-batch of state-action pairs \((\textbf{s}, \textbf{a})\) as

\begin{equation}
\begin{split}
    \mathcal{L}(\boldsymbol{\theta}) = \mathbb{E}\Bigg[ \bigg[ \textbf{r}_t + \gamma Q^\pi\bigg(\mathbf{s}_{t+1} &, \arg \max_{\mathbf{a}\in \boldsymbol{A}}~ Q^\pi(\mathbf{s}_{t+1}, \mathbf{a}_t; \boldsymbol{\theta}); \boldsymbol{\theta}' \bigg) \\
    &-Q^\pi(\mathbf{s}_t,\mathbf{a}_t;\boldsymbol{\theta})\bigg]^2\Bigg],
    \label{eqn:loss_fn}
\end{split}
\end{equation} \normalsize
The Target DDQN's parameter set \(\boldsymbol{\theta}'\) is trained using an experience replay buffer that stores previous experiences \((s, a, r, t)\) to minimize temporal correlations. Once the buffer exceeds a predefined threshold \(\kappa\), mini-batches of experiences are sampled to update the DDQN parameters. 

During training, the objective is to reduce the loss $\mathcal{L}(\boldsymbol{\theta})$ across a mini-batch of state-action pairs, specifically achieving \(\boldsymbol{\hat{\theta}}= \arg \underset{\boldsymbol{\theta}}{\min}~ \mathcal{L}(\boldsymbol{\theta};\textbf{s},\textbf{a})\). The backpropagation algorithm is effectively utilized to compute \(\nabla_{\boldsymbol{\theta}}\mathcal{L}(\boldsymbol{\theta};\textbf{s},\textbf{a})\), which represents the gradient of the loss with respect to the parameters of the DDQN for the given state-action pairs. By applying stochastic gradient descent (SGD) with the calculated gradients \(\nabla_{\boldsymbol{\theta}}\mathcal{L}(\boldsymbol{\theta};\textbf{s},\textbf{a})\), the parameters of the DDQN are updated as $\boldsymbol{\theta} = \boldsymbol{\theta} - \eta\nabla_{\boldsymbol{\theta}}\mathcal{L}(\boldsymbol{\theta};\textbf{s},\textbf{a})$. The learning rate \(0 < \eta < 1\) is adjusted using the adaptive moment estimation (Adam) optimizer for faster computation and a learning rate scheduler to gradually decay the rate for stable learning and efficient convergence.

To enable the CIoT agent to explore the environment, identify optimal strategies, and manage the exploration-exploitation trade-off, we use the upper confidence bound (UCB) algorithm. The UCB algorithm adjusts the Q-values using $\overline{Q}^\pi (s,a) = Q^\pi (s,a) + U_a^t,$ where \(U_a^t\) is the calculated expected reward, defined as

\begin{equation} 
    U_a^t = \hat{r}_a^t + \sqrt{\frac{c' \ln{t}}{C_a^t}},
\end{equation}\normalsize
with \(c'\) being a hyperparameter of the UCB algorithm. The computed expected reward \(U_a^t\) combines the estimated reward \(\hat{r}_a^t\) with an adjustment factor that depends on the number of the time period (frame number * \(T\) + \(t\)) and the number of times action \(a\) has been chosen \(C_a^t\). If action \(a_t\) has been selected \(C_a^t\) times by the end of time slot \(t\) (ranging from 0 to \(t\)), then \(\hat{r}_a^t\) is calculated as 
\begin{align}
    \hat{r}_a^t = \frac{\sum_{i=1}^{C_a^t}r_{a,i}^t}{C_a^t},
\end{align} \normalsize
where \(r_{a,i}^t\) is the reward for action \(a_t\) at the \(i\)th selection. Subsequently, the action chosen by the UCB algorithm is incorporated into the DDQN training, and both the action count \(C_a^t\) and the expected reward \(\hat{r}_a^t\) are updated accordingly. The adjustment of the Q-value using the UCB strategy is illustrated in Fig. \ref{fig:dqn_model}. The training process for the DDQN-UCB algorithm is shown in Algorithm \ref{alg:algorithm1}. \textcolor{black}{The UCB algorithm has a space complexity of \( \mathcal{O}(z) \), where \( z \) is the number of output neurons of the DDQN. Additionally, its computational complexity is \( \mathcal{O}(1) \). The space complexity of the DDQN-UCB is $\mathcal{O}(\mathcal{M})$ and its computational complexity per time step is $\mathcal{O}(\theta)$, where $\theta$ is the total number of parameters in the DDQN.}

\begin{algorithm}[t!]
\caption{The proposed DDQN-UCB Algorithm}\label{alg:algorithm1}
\textbf{Input:} Cognitive IoT environment simulator and parameters.

\textbf{Output:} Optimal action $a_t$ in each time slot $t$.

Initialize experience replay memory $\mathcal{M}$ with size $m$.

Initialize $B_0$, $\eta$, $\gamma$, $\kappa$, and $c'$.

\For{episode= 1,...,episodes}{
  
    \For{t= 1,...,T}{
        Observe the state $s_t$;
        \If{$\mathcal{M}$ is not full}{
        
        Sample a random action $a_t$;
        }

        \Else{ 
            Calculate $U_a^t \leftarrow \hat{r}_a^t\hat+\sqrt{\frac{c'\ln{t}}{C_a^t}}$;

            Adjust Q-value $\overline{Q}^\pi (s,a) \leftarrow Q^\pi (s,a) + U_a^t$;
            
            Get action $a_t$ according to the policy of adjusted Q-value;           
            }

            Get the reward $r_t$ using (\ref{eq:reward}); 
            
            observe the next state $s_{t+1}$ ;
        
            Store $\mathcal{M}\leftarrow(s_t,a_t,r_t,s_{t+1})$;
        
            Update action count, $C_a^t \leftarrow C_a^t + 1$; 

            Update $\hat{r}_a^t \leftarrow (\sum_{i=1}^{C_a^t}r_{a,i}^t)/C_a^t$;
            
            Sample a mini-batch $X$ from $\mathcal{M}$ ;

            Predict Target Q-values using 
            
            $\textbf{r}_t + \gamma \max_{\textbf{a}\in \boldsymbol{\mathcal{A}}}~ Q^\pi(\textbf{s}_{t+1},\textbf{a}_{t+1};\boldsymbol{\theta}')$ ;

            Predict Q-values using $Q^\pi(\boldsymbol{s},\boldsymbol{a};\boldsymbol{\theta})$;

            Calculate the loss in (\ref{eqn:loss_fn});
            
            Update $\boldsymbol{\theta}$ of DDQN online;
            
            \If{episode*t \text{mod} $\kappa$ = 0}{
            Update $\boldsymbol{\theta}'$ of Target DDQN online as $\boldsymbol{\theta}' \leftarrow \boldsymbol{\theta}$ ;
             }

}
         Update $\epsilon$ and the state $s_{t+1} = s_t$;
         
         Update $\eta$ using scheduler ;
}
\end{algorithm}

\section{Simulation Model and Results}
\textcolor{black}{In this section, we analyze the performance of our proposed DRL approach in a CIoT network with a Tx-Rx pair sharing a channel with a PU Tx-Rx pair in a non-line-of-sight (NLoS) environment. We consider a channel noise variance of $\sigma^2=10^{-3}$ and a path-loss exponent of \( \alpha = 4 \). The device distances are \( d_{sp} = d_{ps} = 1.8 \, \text{m} \) and \( d_{ss} = 1.5 \, \text{m} \). Time-slotted transmissions are considered over \( T = 30 \) slots (each lasting \( \tau = 1 \, \text{s} \)), with the PU Tx using \( L = 18 \) slots at \( P_p = 0.2 \, \text{W} \) and an interference threshold of \( I_{th} = 0.1 \, \text{W} \). The CIoT Tx's battery capacity is \( B_{max} = 0.5 \, \text{W} \), with dynamic selection of \( \rho_t \in [0, 0.1, \dots, 1] \) and \( P_s^t \in [0, 0.01, \dots, 0.1] \). The harvested energy at a time slot $t$ follows $e_t \sim \Gamma(0.5,1)$ and the energy efficiency factor is \( \mu = 0.9 \). The DDQN architecture has four layers with \( j = 6 \), \( h_1 = 512 \), \( h_2 = 128 \), and \( z = 121 \) neurons, updating the target network every 200 iterations. Training uses an initial learning rate of \( 2 \times 10^{-4} \), halved every 500 episodes, over 2500 episodes with mini-batches of 80 frames. Constraint violations incur a penalty \( \phi = 7 \). The Target DDQN is updated every 200 iterations. The replay buffer holds \( \kappa = 333 \) experiences, with a discount factor \( \gamma = 0.99 \) and UCB hyperparameter \( c' = 2.5 \).}

\begin{figure*}
    \centering
    \includegraphics[width = 2\columnwidth]{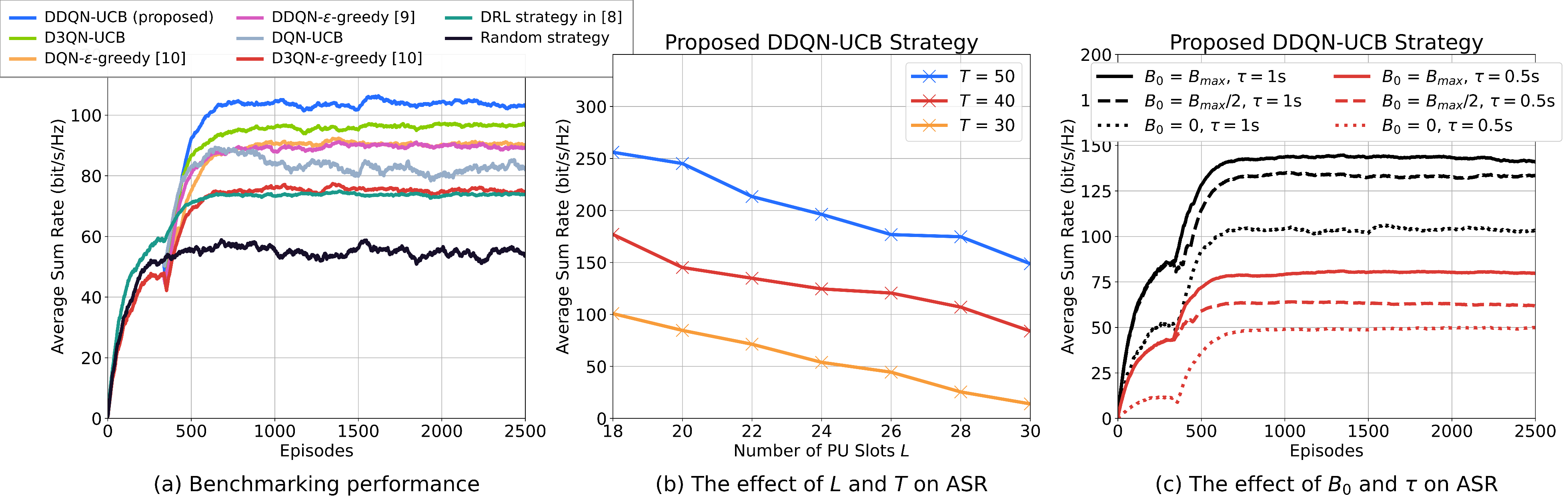}
    \caption{(a) Benchmarking the ASR performance of our DDQN-UCB strategy in comparison to the existing strategies in the literature\cite{tashma_drl, IoT_2024_Nada_Nadia, Zarif_2022_dueling_vanilla_DQN}, (b) Illustrating the effect of varying the number of slots occupied by PU $L$ and the number of time slots $T$ on our proposed DDQN-UCB strategy, and (c) Presenting the impact of varying the initial battery level $B_0$ and the duration of each time slot $\tau$ on our proposed DDQN-UCB strategy.}
    \label{fig:results}
\end{figure*}

In Fig.~\ref{fig:results} (a), we provide a comprehensive comparison of the average sum rate (ASR) achieved by the CIoT agent using our proposed DDQN-UCB strategy against other existing benchmarks. The strategies included in this figure are: the random strategy, where actions are chosen randomly at each time step; the DRL strategy from \cite{tashma_drl}, which assumes a constant TS factor ($\rho_t =0.5$) and a learnable transmission power $P_s^t$; and the DQN and Dueling-DDQN (D3QN) strategies from \cite{Zarif_2022_dueling_vanilla_DQN}. We have applied both $\epsilon$-greedy and UCB methods for balancing the exploration-exploitation trade-off for each of these learning strategies.

Initially, all DRL strategies promote action exploration, which results in penalties as shown in Fig.~\ref{fig:results}(a). However, as training advances, our DDQN-UCB strategy progressively surpasses all presented methodologies. This improvement is attributed to the proposed UCB exploration, which adeptly explores actions that optimize long-term performance, thereby enhancing resource allocation efficiency and increasing ASR. Nevertheless, this enhancement is not solely due to UCB; substituting the DRL framework with DQN or D3QN while using UCB fails to achieve the optimal strategy, underscoring that the DDQN architecture combined with UCB is superior. \textcolor{black}{Moreover, the performance gap between DDQN-UCB and D3QN-UCB clearly indicates that more complex architectures, like D3QN, do not necessarily guarantee better performance, which can be attributed to the additional layers in D3QN.} Limiting the TS factor $\rho$ as per \cite{tashma_drl} and focusing solely on power optimization does not yield the most effective strategy for maximizing performance. Moreover, neglecting to jointly optimize the TS factor and transmission power leads to suboptimal outcomes.

In Fig.~\ref{fig:results} (b), we illustrate how the number of slots occupied by the PU $L$ affects the ASR of the CIoT agent using our proposed DDQN-UCB strategy, comparing results across different numbers of time slots $T$. As shown in the figure, as $L$ increases, the ASR decreases due to the increased limitations on the CIoT agent's actions. With more PUs, the CIoT agent must adhere to the interference threshold in more slots, resulting in a higher likelihood of penalties and a lower data rate of transmission. Additionally, by increasing the number of time slots $T$ in each episode, the ASR increases because the CIoT agent has more opportunities for transmission or harvesting, resulting in a higher ASR.

In Fig.~\ref{fig:results} (c), we present the impact of varying the initial battery level $B_0$ and the duration of each time slot $\tau$ on the ASR achieved by the CIoT agent using our proposed DDQN-UCB strategy. Increasing $B_0$ results in a higher ASR. This can be attributed to the CIoT agent facing fewer penalties, especially in the initial time slots, when it has a higher battery capacity. Additionally, it is evident that as $\tau$ increases, the ASR also rises. This is because longer time slots allow for greater energy accumulation during the EH period $\rho_t\tau$, subsequently enhancing the CIoT agent's capability for data transmission in later slots. Likewise, the period designated for transmission $(1-\rho_t)\tau$ also benefits from a broader window to transmit more data. Overall, the consistent convergence of our UCB-driven DRL strategy across various scenarios indicates its potential to enhance CIoT system performance effectively in diverse settings.

\section{Conclusion}
In this letter, we introduce a novel DDQN-UCB algorithm for joint management of time allocation and power control. Unlike previous approaches, we integrate time switching and transmit power as learnable parameters, enabling a CIoT Tx to dynamically adjust its operations based on real-time environment conditions. By employing a model-free MDP framework and utilizing a DDQN architecture with UCB exploration, our approach effectively balances the exploration-exploitation trade-off, leading to superior long-term performance. \textcolor{black}{Through comprehensive simulations, we demonstrate that our proposed DDQN-UCB strategy not only meets its objectives with notable success over existing benchmarks, but also efficiently adapts to varying conditions, thereby enhancing the robustness and reliability of dynamic CIoT networks. A future direction involves expanding the current spectrum sharing scheme to accommodate larger-scale CIoT networks with multiple PU and SU pairs, addressing the scalability challenges.}

\bibliography{ref.bib}

@INPROCEEDINGS{8885426,  author={Anh, Tran The and Luong, Nguyen Cong and Niyato, Dusit and Liang, Ying-Chang and Kim, Dong In},  booktitle={Proc. IEEE Wireless Commun. Netw. Conf. (WCNC)},   title={{Deep Reinforcement Learning for Time Scheduling in RF-Powered Backscatter Cognitive Radio Networks}},   year={2019},  volume={},  number={},  pages={1-7},  doi={10.1109/WCNC.2019.8885426}}

@ARTICLE{9680720,  author={Guo, Shaoai and Zhao, Xiaohui},  journal={IEEE Trans. Cogn. Commun. Netw.},   title={{Deep Reinforcement Learning Optimal Transmission Algorithm for Cognitive Internet of Things With RF Energy Harvesting}},   year={2022},  volume={8},  number={2},  pages={1216-1227},  doi={10.1109/TCCN.2022.3142727}}

@INPROCEEDINGS{lu_full_duplex_power_control,
  author={Lu, Min and Zhou, Bin and Bu, Zhiyong and Zhao, Yu},
  booktitle={Proc. IEEE Wireless Commun. Netw. Conf. (WCNC)}, 
  title={{A Learning Approach Towards Power Control in Full-Duplex Underlay Cognitive Radio Networks}}, 
  year={2022},
  volume={},
  number={},
  pages={2017-2022},
  doi={10.1109/WCNC51071.2022.9771806}}

@INPROCEEDINGS{tashma_drl,
  author={Tashman, Deemah H. and Cherkaoui, Soumaya and Hamouda, Walaa},
  booktitle={Proc. Int. Wireless Commun. Mobile Comput. Conf. (IWCMC)}, 
  title={{Performance Optimization of Energy-Harvesting Underlay Cognitive Radio Networks Using Reinforcement Learning}}, 
  year={2023},
  volume={},
  number={},
  pages={1160-1165},
  doi={10.1109/IWCMC58020.2023.10182973}}

@ARTICLE{9915473,
  author={Zhang, Renhao and Li, Xuanheng and Zhao, Nan},
  journal={IEEE Trans. Veh. Technol.}, 
  title={{When DSA Meets SWIPT: A Joint Power Allocation and Time Splitting Scheme Based on Multi-Agent Deep Reinforcement Learning}}, 
  year={2023},
  volume={72},
  number={2},
  pages={2740-2744},
  doi={10.1109/TVT.2022.3213243}}

@ARTICLE{IoT_2024_Nada_Nadia,
  author={Khalek, Nada Abdel and Abdolkhani, Nadia and Hamouda, Walaa},
  journal={IEEE Internet Things J.}, 
  title={{Deep Reinforcement Learning for Joint Power Control and Access Coordination in Energy Harvesting CIoT}}, 
  year={2024},
  volume={11},
  number={19},
  pages={30833-30846},
  doi={10.1109/JIOT.2024.3416371}}

@ARTICLE{Zarif_2022_dueling_vanilla_DQN,
  author={Hossein Zarif, Amir and Azmi, Paeiz and Yamchi, Nader Mokari and Javana, Mohammad Reza and Jorswieck, Eduard A.},
  journal={IEEE Trans. Green Commun. and Netw.}, 
  title={{AoI Minimization in Energy Harvesting and Spectrum Sharing Enabled 6G Networks}}, 
  year={2022},
  volume={6},
  number={4},
  pages={2043-2054},
  doi={10.1109/TGCN.2022.3186279}}

@ARTICLE{nada_survey_2023,
  author={Khalek, Nada Abdel and Tashman, Deemah H. and Hamouda, Walaa},
  journal={IEEE Commun. Surv. Tutor.}, 
  title={{Advances in Machine Learning-Driven Cognitive Radio for Wireless Networks: A Survey}}, 
  year={2024},
  volume={26},
  number={2},
  pages={1201-1237},
  doi={10.1109/COMST.2023.3345796}}

@ARTICLE{R2,
  author={Omidkar, Atefeh and Khalili, Ata and Nguyen, Ha H. and Shafiei, Hossein},
  journal={\textcolor{black}{IEEE Internet Things J.}}, 
  title={\textcolor{black}{Reinforcement-Learning-Based Resource Allocation for Energy-Harvesting-Aided D2D Communications in IoT Networks}}, 
  year={\textcolor{black}{2022}},
  volume={\textcolor{black}{9}},
  number={\textcolor{black}{17}},
  pages={\textcolor{black}{16521-16531}},
 }

@ARTICLE{10673973,
  author={Abdolkhani, Nadia and Khalek, Nada Abdel and Hamouda, Walaa},
  journal={IEEE Internet Things J.}, 
  title={{Deep Reinforcement Learning for EH-Enabled Cognitive-IoT Under Jamming Attacks}}, 
  year={2024},
  volume={11},
  number={24},
  pages={40800-40813},
  doi={10.1109/JIOT.2024.3457012}}

@INPROCEEDINGS{deap_learning,
  author={Khalek, Nada Abdel and Hamouda, Walaa},
  booktitle={Proc. IEEE Glob. Commun. Conf. (GLOBECOM)}, 
  title={{DEAP Learning: A Data-Driven Approach to Unsupervised Cooperative Spectrum Sensing}}, 
  year={2023},
  volume={},
  number={},
  pages={6389-6394},
  doi={10.1109/GLOBECOM54140.2023.10437464}}
\bibliographystyle{IEEEtran}

\end{document}